\documentclass[aps,prl,showpacs,twocolumn]{revtex4}
\usepackage{graphicx}
\usepackage{amssymb}
\usepackage{amsmath}
\usepackage{color}
\usepackage{dsfont}
\usepackage{paralist}
\usepackage{bm}
\usepackage{verbatim}

\def\be{\begin{equation}}
\def\ee{\end{equation}}
\def\bea{\begin{eqnarray}}
\def\eea{\end{eqnarray}}

\def\rf#1{(\ref{#1})}

\def\rf#1{(\ref{#1})}

\def\rfs#1{Eq.~\rf{#1}}

\begin{document}

\title{Hall viscosity as a topological invariant}


\author{V.~Gurarie} 
\affiliation{Physics Department, University of Colorado, Boulder, CO 80309, USA}

\begin{abstract}
	
  {Hall conductance of noninteracting fermions filling a certain number of Landau levels can be written as a  topological invariant. A particular version of this invariant 
 when expressed in terms of the single particle Green's functions directly generalizes to cases when interactions are present including those of fractional Hall states, although in those cases this invariant no longer corresponds to Hall conductance. We argue that when evaluated for fractional Hall states this invariant gives twice the total orbital spin
  of fermions which in turn is closely related to 
 the Hall viscosity, a quantity characterizing the integer and fractional Hall states which recently received substantial attention in the literature. }
\end{abstract}



\date{\today}
\maketitle

Energy levels of a particle moving in a two dimensional plane in a magnetic field perpendicular to that plane form bands called Landau levels. If all the levels lying below
a certain fixed energy located in  a band gap (Fermi energy) are filled by non-interacting fermions, we obtain what can be called the integer quantum Hall system. The conductance of this system
when written in units of $e^2/h$, is integer valued and can be thought of as a topological invariant, the sum of Chern numbers of the filled bands \cite{Thouless1982}.

It is possible to re-express this conductance as a certain topological winding of the fermionic Green's function. Indeed, the Green's function of this system of fermions is $G_{ab}(\omega,{\bf k})$, where $\omega$ is frequency conjugate to imaginary time, ${\bf k}$ is the two dimensional (quasi-)momentum, and indices $a$, $b$ refer to the Landau levels. It can be mathematically thought of as a matrix valued function of frequency $\omega$ and quasimomentum ${\bf k}$. Such matrix valued functions are known to fall into equivalence classes such that a Green's function belonging to one class cannot be smoothly deformed into a Green's function belonging to a different class (in mathematics those are usually called the homotopy classes). These classes can be labelled by integers. It has been known for quite some time that these integers coincide with the Hall conductance expressed in units of $e^2/h$, in other words with combined Chern numbers of filled bands \cite{Niu1985}. The fact that the integer Hall system has an excitation gap implies that the Green's function is never infinity and is a smooth function of frequency and momenta, which is necessary for these classes to be well-defined.

Consider now a system of interacting fermions moving in the same two dimensional plane in a perpendicular magnetic field. Depending on their density, they can still be in an integer quantum Hall phase, but they also can be in a fractional Hall state, or in some cases in gapless states. As long as the system has an excitation gap (that is, the Green's function is not singular), and as long as the Green's function never vanishes for any frequency and momenta, it is still a smooth function of its arguments and is also labelled by integers representing the homotopy classes discussed earlier \cite{Gurarie2011}. But these integers no longer necessarily coincide with the Hall conductance. First of all, the expression for the Hall conductance in terms  of the single-particle fermionic Green's function is derived in the absence of any interactions, and the derivation clearly breaks down once the interactions among fermions are turned on. Moreover, the Hall conductance in fractional Hall states when expressed in terms of $e^2/h$ is a non-integer fraction, and cannot be directly related to the integers labeling homotopy classes of Green's functions. Establishing the meaning of those integers in a fractional Hall regime is therefore not entirely trivial. In Ref.~\cite{Essin2013} these integers were calculated for certain fractional Hall states, but no conclusions were drawn as to their meaning.

We would like to argue that these integers, when properly defined as explained below, give twice the total orbital spin of the fermion \cite{Zee1992a}. In turn, 
this spin is known to  
coincide with the shift ${\cal S}$ of the fractional quantum Hall state \cite{Zee1992a} in some although not all cases \cite{Read2009}.  Additionally, the shift is known to represent the Hall viscosity \cite{Avron1995}, a certain non-dissipative transport coefficient $\eta$ which received significant attention in the literature recently, via the relation
\be  \label{eq:visc} \eta = \frac{\hbar n {\cal S}}{4},
\ee  where $n$ is the fermion number density \cite{Vignale2009,Read2009,Haldane2009,Rezayi2011,Son2012}.
Taken together, this means that the Hall viscosity  is often, although not always, given by a topological invariant, which is the main result of this paper. 

More precisely, simple arguments \cite{Read2009} can be given that the total orbital spin of the fermion coincides with the shift 
for all the states whose wave functions are given by the so-called conformal blocks of relevant conformal field theories \cite{Read1991}, including
the Laughlin states, the Read-Moore (Pfaffian) state, as well
as Read-Rezayi (parafermion) states \cite{Rezayi1999}. 
In other cases, including the Abelian hierarchy states and 
some of the states obtained from the Read-Rezayi  states by a particle-hole transformation the shift might not be equal to the spin of the fermion (and might not even be an integer), instead
being represented by an ``average" spin as opposed to the total spin \cite{Zee1992a, Read2009}. In those cases, the direct relationship between the invariant and the Hall viscosity breaks down, although a slightly more indirect relationship remains. 

Let us now present the derivation of this result. 
A Green's function of a system of elections moving in a two dimensional lattice without boundaries (for example, with periodic boundary conditions) can be defined as
\be \label{eq:greendef} G_{ab}(\omega, {\bf k}) = -i \int d\tau e^{i \omega \tau} \, \left\langle {\cal T} \, \hat a_a(\tau, {\bf k}) \, \hat a^\dagger_b(0, {\bf k}) \right\rangle.
\ee Here $\hat a$ and $\hat a^\dagger$ are annihilation and creation operators of fermions in bands labelled by the subscript of these operators at an appropriate imaginary time and at appropriate momenta and ${\cal T}$ is the usual imaginary time ordering operation \cite{AGDBook}. 
In terms of this Green's function the integer-valued topological invariant can be written in the following form \cite{Niu1985}
\be  \label{eq:invariant} N = \sum_{\alpha, \beta, \gamma} \epsilon_{\alpha \beta \gamma}  \, {\rm tr} \, \int \frac{d\omega d^2 k}{24 \pi^2} G^{-1} \partial_\alpha G G^{-1} \partial_\beta G G^{-1} \partial_\gamma G,
\ee
where $G_{ab}^{-1}(\omega, {\bf k})$ is a matrix inverse to $G_{ab}(\omega, {\bf k})$, the variables $\alpha$, $\beta$ and $\gamma$ take values $0$, $1$ and $2$, and $\partial_0 = \partial/\partial \omega$, $\partial_1 = \partial/\partial {k_x}$, $\partial_2 = \partial/\partial {k_y}$.

\begin{figure}[htbp]
	\centering
	\includegraphics[width=0.3\textwidth]{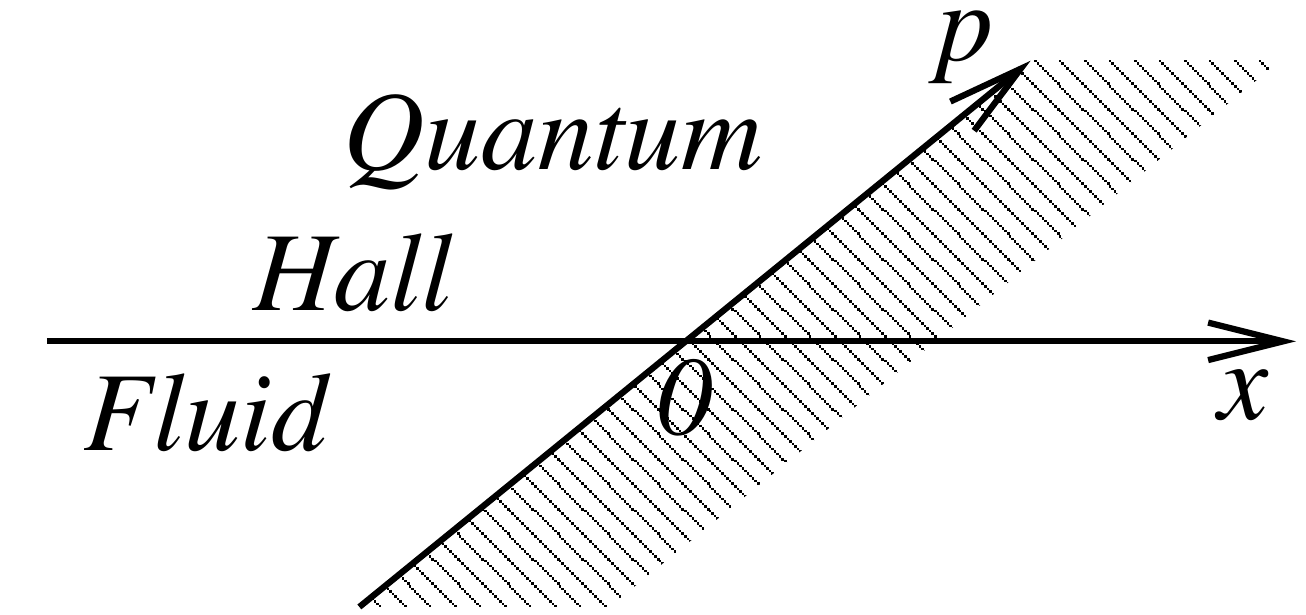}
	\caption{\label{boundary}
	A two dimensional half-plane geometry with a boundary perpendicular to the $x$-axis. Quantum Hall fluid is confined to $x<0$. The axis perpendicular to $x$ is labeled by $p$, the momentum along the $y$-axis.	}
\end{figure}

\begin{figure}[htbp]
	\centering
	\includegraphics[width=0.15\textwidth]{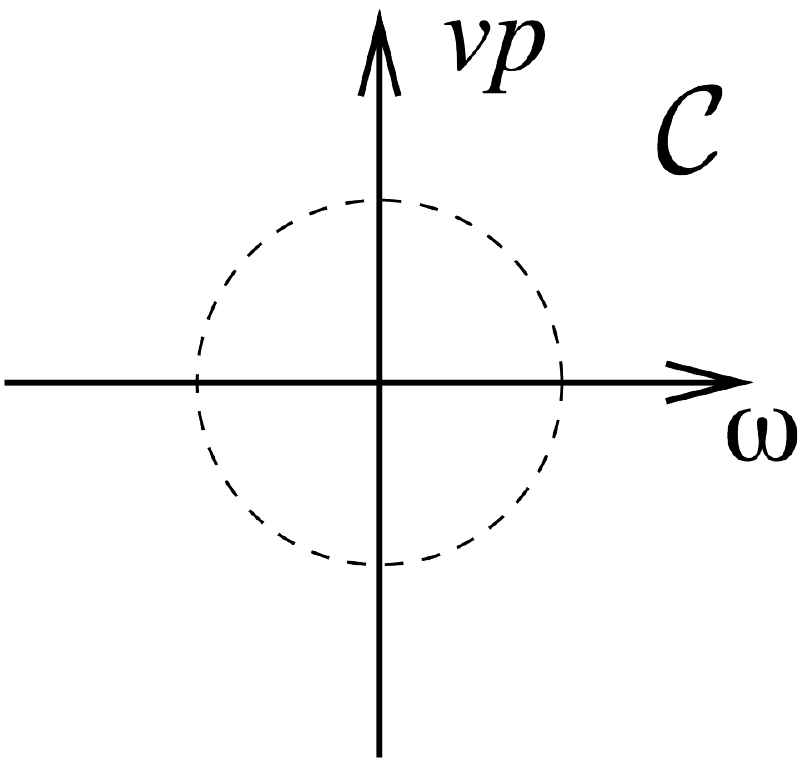}
	\caption{\label{contour}
	A contour of integration ${\cal C}$ in Eq.~\rf{eq:nedge} where $v$ is some arbitrarily chosen velocity.	}
\end{figure}

While in the absence of interactions this invariant can be shown to be equal to Hall conductance up to a factor of $e^2/h$ via the application of the Kubo formula, and while
even with interactions this expression produces an integer, the general physical meaning of $N$ can be established in the following way. Consider a half-plane geometry
where a two dimensional system extends to $x<0$ and terminates at $x=0$ as depicted in Fig.~\ref{boundary}. 
In the presence of the boundary, the Green's function of this system is given by
\be G^B_{ab}(\omega,p; x,x') =-i \int d\tau e^{i \omega \tau} \left\langle {\cal T} \, \hat a_a(\tau, p; x) \, \hat a_b^\dagger(0, p; x') \right\rangle,
\ee
where $\hat a(\tau,p; x)$ annihilates a fermion with momentum $p$ along the boundary at the position $x$ perpendicular to the boundery and at imaginary time $\tau$. Consider the following expression
\begin{eqnarray} \label{eq:nedge} && N_{\rm edge} = \\ &&\sum_{a,b,\mu} \oint_{\cal C} \frac{dk^\mu}{2\pi i} \int dx dx' {G^B_{ab}}^{-1}( \omega, p; x, x') \partial_\mu {G^B_{ba}}(\omega,p; x', x). \nonumber
\end{eqnarray}
Here $k^0 = \omega$, $k^1=p$, and the integral proceeds over a contour ${\cal C}$ which is a circle in the $\omega$-$p$ plane as shown in Fig.~\ref{contour}.  We will see below that
$N_{\rm edge}$ is also an integer-valued topological invariant. 

The following crucial for our purposes result has been established in Refs~\cite{VolovikBook1,EssinGurarie2011}. 
Define the bulk Green's function by the following Wigner transform
\be G(\omega, {\bf k}) = \lim_{R \rightarrow -\infty} \int dr \, e^{-i r k_x} G^B \left(\omega,k_y; R+\frac r 2, R- \frac r 2 \right).  \label{eq:wigner} 
\ee
The limit $R \rightarrow -\infty$ is needed to move far away from the boundary. It is natural to assume $G(\omega, {\bf k})$ defined in this way coincides with the Green's function of the system without boundaries. Using $G(\omega,{\bf k})$ found in this way we can construct the invariant $N$ as in Eq.~\rf{eq:invariant}. It can be shown that $N$ from Eq.~\rf{eq:invariant} and $N_{\rm edge}$ from Eq.~\rf{eq:nedge} are always equal to each other, as long as Eq.~\rf{eq:wigner} is used to relate $G^B$ to $G$. 

This allows us to calculate $N$ by calculating $N_{\rm edge}$ instead. Even though $N_{\rm edge}$ appears to be more complicated than $N$, in fact it will be straightforward to evaluate in a general fractional Hall state. To do that, we first introduce the eigenvectors and eigenvalues of the Green's function in the presence of the boundary $G^B$,
\be  \label{eq:ev} \sum_b \int dx' G^B_{ab}(\omega, p; x, x')\, \psi^{(n)}_b(x') = g_n(\omega,p)  \, \psi_a^{(n)} (x).
\ee
$g_n(\omega,p)$ can be thought of as Green's functions of modes propagating along the boundary. Indeed, it is straightforward to see that we can define an annihilation operator of the boundary mode
\be \hat a_n(\tau,p) =\sum_a \int dx \, {\psi^*_a}^{(n)} (x) \, \hat a_a(\tau, p; x).
\ee
With its help, 
Eq.~\rf{eq:ev} implies that
\be g_n(\omega,p) =-i  \int d\tau e^{i \omega \tau} \left\langle {\cal T} \, \hat a_n(\tau,p) \, \hat a_n^\dagger(0,p) \right\rangle,
\ee
that is, $g_n$ are indeed the Green's functions of modes labelled by their momentum along the boundary. 

Now Eq.~\rf{eq:nedge} can be rewritten in the following way
\be \label{eq:gedge} N_{\rm edge} = \sum_n \oint_{\cal C} \frac{dk^\mu}{2\pi i} \, \partial_\mu \ln g_n(\omega,p).
\ee
This is nothing but  a sum of phases accumulated by $g_n$ over a circular contour in the $\omega-p$ plane, normalized by $2\pi i$. This will necessarily be an integer, justifying the identification of $N_{\rm edge}$ with a topological invariant. Now among $g_n$ most describe fermions moving far away from the boundary. 
However some of those $g_n$ correspond to the gapless edge modes.

To elucidate this further, let us examine the contribution to $N_{\rm edge}$ coming from typical boundary modes encountered in a quantum Hall fluid \cite{Wen1990}. A simple integer Hall boundary mode with the dispersion $vp$ where $v$ is the velocity of the mode  has a Green's function $g = 1/(i \omega - v p)$. The phase it accumulates over a circular contour in the $\omega-p$ plane when normalized by $2 \pi i$ is simply equal to $1$. A more complicated fractional Hall mode, for example in a Laughlin state with a filling fraction $\nu = 1/(2m+1)$ where $m$ is some positive integer has a boundary fermion Green's function which is easiest to find in the coordinate-imaginary time domain
$g(\tau,x) \sim 1/{(x - i v \tau)^{2m+1}}$. However for its use in Eq.~\rf{eq:gedge} we need to know it in the frequency-momentum domain. The appropriate Fourier transform is not straightforward as the Green's function has a strong UV singularity which needs to be regularized. A correct regularization was discussed in the literature on Luttinger liquids (\cite{GiamarchiBook}; see also \cite{KunYang2004}) and involves integrating over $x$ first and then over $\tau$
\be \label{eq:conventional} g(\omega,p) = \int d\tau e^{i \omega \tau} \int   \frac{dx \, e^{i p x}}{(x - i v \tau)^{2 m+1}} \sim \frac{p^{2m}}{i \omega - v p}.
\ee 
This Green's function does not have a well defined phase as $p$ is taken to zero. Eq.~\rf{eq:gedge} with it is not well defined. Instead, we could define a Green's function with a rotationally invariant cutoff \cite{Wen1990}
\be \label{eq:rotational} g_r(\omega,p) = \int dx d\tau    \frac{  f(x^2+v^2 \tau^2)  \, e^{i (\omega \tau + p x)}}{(x - i v \tau)^{2 m+1}}   \sim  \frac{(i \omega+ v p)^{2m}}{i \omega - v p}.
\ee Here $f$ is some function whose precise form is unimportant, but which quickly goes to 1 if  its argument exceeds some arbitrarily chosen small cutoff $a^2$ and goes
to zero sufficiently quickly as its argument goes to zero to make the integral convergent, and the right hand side of this equation is valid for $\omega/v$ and $p$ much smaller than $1/a$. 
Unlike $g$ from Eq.~\rf{eq:conventional}, $g_r$ from Eq.~\rf{eq:rotational} has a well defined phase. When substituted for $g_n$ in Eq.~\rf{eq:gedge}, $g_r$ produces the contribution of $2m+1$. Note however that unlike the more physical function $g(\omega,p)$, $g_r(\omega,p)$ is not a conventional Green's function. For example, it grows at large frequency while a conventional Green's function must always decay as $1/\omega$ \cite{AGDBook}. 

More generally, borrowing the notations from conformal field theory applicable at the boundary \cite{Read1991}, we can imagine a boundary Green's function defined by a boundary fermion operator with holomorphic and antiholomorphic dimensions $\Delta$, $\bar \Delta$,
\be g \sim \frac{1}{(x  - i v \tau)^{2 \Delta} (x+i v \tau)^{2 \bar \Delta}}.
\ee
While the physical Fourier transform along the lines of Eq.~\rf{eq:conventional} produces the Green's function in frequency-momentum space without a well defined phase, just as in the simpler example of Eq.~\rf{eq:conventional}, the phase if rotationally invariant cutoff is employed
\be \label{eq:cirri} g_r = \int  dx d\tau  \frac{ f(x^2+v^2 \tau^2)  \, e^{i (\omega \tau + p x)}}{(x - i v \tau)^{2 \Delta} (x+i v \tau)^{2 \bar \Delta}}
\ee
is clearly well defined. Its accumulation over a circular contour, which can be read directly off Eq.~\rf{eq:cirri} without having to do the integral in it explicitly thanks to its rotational invariance, is $2( \Delta- \bar \Delta)$, which is twice the so-called conformal spin of this fermion boundary operator.   
The message coming from these considerations indicate that we need to consider Green's function with a rotationally invariant cutoff. Note that an even more general Green's function would include several branches of excitations at different velocities. Arguments can be given however that the Fourier transform with a rotationally invariant cutoff with an arbitrary velocity $v$ which does not have to match any of the actual physical velocities of the edge excitations still produces the Green's function with the phase accumulating according to the conformal spin of the boundary operator \cite{Essin2013}. 

Let us also note that bulk modes (the ones which propagate far from the boundary) do not contribute at all to $N_{\rm edge}$. The Green's functions for these modes have a form $g \sim 1/(i \omega - \epsilon(p))$ where $\epsilon(p)$ never changes sign as a function of $p$ to maintain the bulk gap. Such function does not wind at all along a circular contour in the frequency-momentum domain (it phase accumulation along this contour is zero). 
The end result of these arguments is that $N_{\rm edge}$ calculated with ${g_r}_n$ when summed over $n$ gives twice the total conformal spin of the fermion boundary operator summed over all the boundary modes. 

To be able to relate $N_{\rm edge}$ to $N$ computed in the bulk, we imagine applying a rotationally invariant cutoff to the functions $g_n(\omega,p)$. This can be accomplished by Fourier transforming them into the euclidean time-space domain and then transforming them back into the frequency-momentum domain while employing the rotationally invariant cutoff. Together the relevant transformation looks like
\be {g_r}_n (\omega,p) = \int d\omega' dp' K(\omega-\omega', p-p') g_n(\omega,p),
\ee where $K$ is the Fourier transform of $f$. 
We can then do it systematically for every $n$ and construct $G^B_r(\omega,p; x,x')$ from $G^B(\omega,p;x,x')$ according to 
\be G^B_r(\omega,p; x,x') = \sum_n {\psi^*_a}^{(n)}(x) {g_r}_n(\omega,p) \psi^{(n)}_b(x'),
\ee
where $\psi^{(n)}$ are the eigenvectors defined in Eq.~\rf{eq:ev}.
This $G^B_r$ satisfies the bulk boundary correspondence in the same way as $G^B$. This implies that $N_{\rm edge}$ calculated with ${g_r}_n$ substituted for $g_n$ in Eq.~\rf{eq:gedge} is equal to $N$ calculated with $G_r$ substituted for $G$ in Eq.~\rf{eq:invariant}. In turn, $G_r$ can directly be constructed from the bulk Green's function according to
\be \label{eq:gr} G_r(\omega,k_x,k_y) = \int d\omega' dk_y' K(\omega-\omega', k_y-k_y') G(\omega',k_x,k_y').
\ee
Defined in this way, $N$ computed in the bulk therefore gives twice the total conformal spin of the fermion boundary operator, as promised earlier. In turn, the 
total conformal spin coincides with the sum of all the components of the  spin vector as defined  in Ref.~\cite{Zee1992a}. This observation also matches the explicit evaluation of
the invariant in Ref.~\cite{Essin2013}.

A practical application of this result is that in the exact diagonalization studies it is now possible to compute this invariant numerically to help identifying the approximately found ground states with the possible candidate states. One difficulty one needs to overcome is that its direct evaluation according to the Eq.~\rf{eq:invariant} may not be easy. Indeed, it involves the knowledge of the entire matrix-valued $G(\omega,{\bf k})$ including $G$ computed in higher Landau levels. In a typical fractional Hall study, we examine fermions confined to a lowest Landau level, and evaluating the entire Green's function in all Landau levels for subsequent substitution into Eq.~\rf{eq:invariant} may be problematic. Fortunately, just like in case of integer Hall effect when the evaluation of Eq.~\rf{eq:invariant} can be replaced by the evaluation of the Chern number of the filled bands, in case of fractional Hall effect in the lowest Landau level it is possible to reduce Eq.~\rf{eq:invariant} to the study of the phase winding of the Green's function in the lowest Landau level in the following way. 

We imagine studying a fractional Hall state
employing basis wave functions of Landau levels with periodic boundary conditions, satisfying
\begin{eqnarray} \Psi_n(x+\ell,y; k_x,k_y) &=& e^{2 \pi i y/\ell+i k_x \ell} \Psi_n(x,y;k_x,k_y), \cr \Psi_n(x,y+\ell; k_x,k_y) &=& e^{i k_y \ell} \Psi_n(x,y;k_x,k_y),
\end{eqnarray}
where $\ell$ is the magnetic length and $n$ labels Landau levels. These functions can be easily found explicitly  \cite{Rezayi1985}. Similar functions can also be found in case if the motion occurs on a lattice. 
In the basis of these functions the Green's function of a fractional Hall state confined to the lowest Landau level (that is, neglecting Landau level mixing justified at sufficiently weak interactions) is essentially diagonal. Its lowest diagonal entry $G_{LLL}(\omega, {\bf k}) = G_{00}(\omega,{\bf k})$ is computed entirely in the lowest Landau level 
according to Eq.~\rf{eq:greendef} with $a=b=0$. Its higher diagonal entries look essentially like non-interacting Green's function $G_{nn}=1/(i \omega - \epsilon_n)$ for $n>0$, where $\epsilon_n>0$ are the energy of the corresponding Landau levels. We then evaluate \rf{eq:invariant} by calculating the trace in the basis of functions $\Psi_n$. The end result of this calculation \cite{Supp} is that one needs to evaluate the phase winding of the Green's function
\be \label{eq:phasewinding} W = \int_{-\infty}^\infty \frac{d\omega}{\pi i} \,  \partial_\omega \ln G_{LLL}(\omega, {\bf k})
\ee at arbitrary ${\bf k}$. In terms of this number, as long as it is odd, the invariant $N$ is given by
\be \label{eq:invwind} N = \frac{W+1}{2} C,
\ee where $C$ is the Chern number of the lowest Landau level. For a particle moving in continuous 2D space in a magnetic field (as opposed to moving in a lattice), this Chern number is simply $C=1$. Note that in case of integer Hall effect in the absence of interactions $G_{LLL} = 1/(i \omega + |\epsilon_0|)$ where $\epsilon_0<0$ is the energy of the lowest Landau level and $W=1$, giving $N=C$ as expected (it has to coincide with the Chern number of the lowest Landau level). 

Therefore, a practical numerical evaluation of $N$ may proceed with the following steps. First the Green's function is computed numerically by exact diagonalization and spectral decomposition \cite{AGDBook} entirely within the lowest Landau level 
\be \label{eq:lehmann} G_{LLL} = \sum_m \frac{ \left| \left\langle m \right| \hat a^\dagger_0 ({\bf k}) \left| {0} \right\rangle  \right|^2 }{i \omega - E_m+E_0} + \sum_m \frac{ \left| \left\langle m \right| \hat a_0 ({\bf k}) \left| {0} \right\rangle  \right|^2 }{i \omega + E_m-E_0}.
\ee
Here $\hat a_0^\dagger({\bf k})$ creates a particle in the lowest Landau level with momentum ${\bf k}$ (in the state $\Psi_0({\bf k})$), $\left|0 \right>$ represents an exact fractional Hall state, and $E_m$ are the energies of exact eigenstates of the interaction Hamiltonian containing one more or one less particle then the fractional Hall state under study. Given $G_{LLL}$ the operation
Eq.~\rf{eq:gr} has to be applied to it.  Finally, its phase winding $W$ has to be determined by employing Eq.~\rf{eq:phasewinding} or  by a direct examination of its phase. Note that
 the expression \rfs{eq:lehmann} by itself cannot give winding different from $W=0$ or $W=\pm 1$. This is easiest to see if one notes that Im~$G=0$ implies $\omega=0$, thus as $\omega$ is taken from $-\infty$ to $+\infty$ the total phase accumulation of $G$ cannot exceed $\pi$. Therefore, the operation Eq.~\rf{eq:gr} is a crucial step to convert $G$ into
 $G_r$ whose phase can wind multiple number of times.  The winding $W$ gives the invariant $N$ according to Eq.~\rf{eq:invwind} (see \cite{Supp} in case if $W$ is not odd integer). 

Finally in recent work it has been argued that the Hall viscosity is closely related to Hall conductivity at finite momentum \cite{Son2012,Read2012}. It would be interesting to see whether this sheds further light on the meaning of the topological invariant Eq.~\rf{eq:invariant}.

{\it Acknowledgements.} This work has been supported by the NSF grants  
DMR-1205303  and PHY-1211914. The author is grateful to A. Essin, P. Ostrovsky, T. Giamarchi, N. Cooper and N. Read for useful conversations at various stages of this project.


\bibliographystyle{apsrev}
\bibliography{FQHE}


\section*{Supplemental material\label{Sec:App}}

\renewcommand{\theequation}{S\arabic{equation}}
\renewcommand{\thefigure}{S\arabic{figure}}
\renewcommand{\thetable}{S\arabic{table}}
\renewcommand{\thetable}{S\arabic{table}}
\renewcommand{\bibnumfmt}[1]{[S#1]}
\renewcommand{\citenumfont}[1]{S#1}

\setcounter{equation}{0}
\setcounter{figure}{0}
\setcounter{enumiv}{0}

\subsection*{Evaluation of the invariant}

We begin with the expression Eq.~(2) for the topological invariant, which we rewrite for convenience as 
\be  \label{eq:invariant1} N = \sum_{\alpha, \beta =1,2 } \epsilon_{\alpha \beta}  \, {\rm tr} \, \int \frac{d\omega d^2 k}{8 \pi^2} G^{-1} \partial_\omega G G^{-1} \partial_\alpha G G^{-1} \partial_\beta G.
\ee  Here $\partial_1 = \partial_{k_x}$, $\partial_2 = \partial_{k_y}$.  
We emphasize that what we use here is not the Green's function itself, but the Green's function $G_r$ transformed with the operation Eq.~(17), but to avoid cluttering equations we drop the subscript $r$ everywhere.

We study a fractional Hall state in the lowest Landau level, in the regime where Landau level mixing can be neglected. In this regime the Green's function is diagonal in the space of Landau levels. This means its eigenfunctions coincide with $\Psi_n$ defined in Eq.~(18). As explained in the main body of the paper, in the basis of these eigenfunctions the Green's function in the excited Landau levels can be thought of as simply the Green's functions of noninteracting electrons,
\be \label{eq:highlevels} g_n=G_{nn} = \frac{1}{i\omega- \epsilon_n}, \ n = 1, 2, \dots
\ee where $\epsilon_n>0$ are the single-particle energy of the excited Landau levels (note that the operation Eq.~(17) does not change $g_n$ because its Fourier transform is UV convergent). The lowest Landau level Green's function is a nontrivial function of frequency and momenta, $g_0=G_{00}(\omega,{\bf k})$, transformed with Eq~(17), which encodes in itself the information about the fractional Hall state.  

We explicitly evaluate the trace in Eq.~\rf{eq:invariant1} to find
\be \int \frac{d\omega d^2k}{8 \pi^2} \sum_{n,m}  \sum_{\alpha \beta} \epsilon_{\alpha \beta}  \left[ \frac{\partial_\omega g_n}{g_n^2 g_m} \left< n \right| \partial_\alpha G \left| m \right> \left< m \right| \partial_\beta G \left| n \right>  \right] \ee
Here the summations over $n$ and $m$ go over $0$, $1$, $2$, $\dots$. The Dirac notations here imply computing the expectation of the matrix $\partial G$ with respect to the basis defined in Eq.~(18). Note that while $G$ itself is diagonal in this basis, its derivatives are not. 

In turn, this can be recast in the following form
\be \int \frac{d\omega d^2k}{16 \pi^2} \sum_{n \not = m}  \sum_{\alpha \beta} \epsilon_{\alpha \beta}  \left[ \frac{\partial_\omega \ln \frac {g_n}{g_m}}{g_n g_m} \left< n \right| \partial_\alpha G \left| m \right> \left< m \right| \partial_\beta G \left| n \right>  \right] \ee
Now we can take advantage of the standard relation
\be \left< n \right| \partial_\alpha G \left| m \right>  = \delta_{nm} \partial_\alpha g  + (g_m-g_n) \left< n \right| \partial_\alpha \left| m \right>.
\ee
Substituting it gives
\be \int \frac{d\omega d^2 k}{16 \pi^2} \sum_{n \not = m} \sum_{\alpha \beta} \epsilon_{\alpha \beta} \frac{(g_n-g_m)^2 \partial_\omega \ln \frac{g_m}{g_n}}{g_n g_m}
\left< n \right| \partial_\alpha \left| m \right> \left< m \right| \partial_\beta \left| n \right>.
\ee
Introduce a variable $ z_{mn}  = g_m/g_n$.
In terms of this variable we find
\begin{eqnarray} \label{eq:con} \sum_{n \not = m}  \sum_{\alpha \beta}   \epsilon_{\alpha \beta} &&  \oint dz_{mn} \int  \frac{ d^2 k}{16 \pi^2}   \left( 1- \frac{2}{z_{mn}} +\frac{1}{z_{mn}^2} \right) 
\times \cr &&
\left< m \right| \partial_\alpha \left| n \right> \left< n \right| \partial_\beta \left| m \right>.
\end{eqnarray}
As $\omega$ is taken from $-\infty$ to $+\infty$, the variable $z_{nm}$ goes over a certain contour in the complex plane which is being integrated over in Eq.~\rf{eq:con}. If both $n$ and $m$ are larger than 0, thanks to the explicit form Eq.~\rf{eq:highlevels}, $z_{mn}$ does not circle about zero and the integral over it is zero (since the only singularities in Eq.~\rf{eq:con} are at $z_{mn}=0$). Therefore, either $n$ or $m$ are zero. It is sufficient to restrict just one of them to zero, to find
\begin{eqnarray} \sum_{n >0}  \sum_{\alpha \beta}   \epsilon_{\alpha \beta} &&  \oint dz_{n} \int  \frac{ d^2 k}{8 \pi^2}   \left( 1- \frac{2}{z_{n}} +\frac{1}{z_{n}^2} \right) 
\times \cr &&
\left< n \right| \partial_\alpha \left| 0 \right> \left< 0 \right| \partial_\beta \left| n \right>,
\end{eqnarray}
where $z_n=g_n/g_0$. Now $z_n$ winds about zero $(W+1)/2$ times, where $W$ was defined in Eq.~(19). Thus the integral over $z_n$ is $\pi i (W+1)$ times the
residue of the pole at $z_n=0$. This gives
\be (W+1) \sum_{n \ge 0} \sum_{\alpha \beta} \int \frac{d^2 k}{4 \pi i}\left< 0 \right| \partial_\alpha \left| n \right> \left< n \right| \partial_\beta \left| 0 \right>.
\ee
Here the sum could be extended over all $n \ge 0$ as the expression above at $n=0$ is zero. 
Finally, this expression up to a factor of $(W+1)/2$ is the Chern number $C$ of the lowest Landau level, Eq.~(5) from Ref.~[1]. Therefore, the answer is
\be N = \frac{W+1}{2} C.
\ee

\end{document}